\documentstyle[12pt,draft,a4,epsf]{article}

\newcommand{\be}{\begin{equation}}
\newcommand{\ee}{\end{equation}}
\newcommand{\bea}{\begin{eqnarray}}
\newcommand{\eea}{\end{eqnarray}}
\newcommand{\th}{{\tilde{h}}}
\newcommand{\tm}{{\tilde{m}}}
\newcommand{\tE}{{\tilde{E}}}
\newcommand{\tP}{{\tilde{P}}}
\newcommand{\tT}{{\tilde{T}}}
\newcommand{\tgb}{{\tilde{\beta}}}

\newcommand{\Real}{\mbox{Re}}
\newcommand{\Imag}{\mbox{Im}}
\newcommand{\sReal}{\mbox{\footnotesize Re}}

\newcommand{\mybibitem}[3]{\bibitem{#2}}

\begin{document}

\title{Optimized Monte Carlo Methods}
\author{Enzo Marinari\\
Dipartimento di Fisica,
Universit\`a di Cagliari,\\
via Ospedale 72,
09100 Cagliari (Italy)\\
}

\maketitle
\begin{abstract}
I discuss optimized data analysis and Monte Carlo methods.  
Reweighting methods are discussed through examples, like Lee-Yang 
zeroes in the Ising model and the absence of deconfinement in QCD.
I discuss reweighted data analysis and multi-hystogramming. 
I introduce Simulated Tempering, and as an example its application
to the Random Field Ising Model. I illustrate Parallel Tempering, and 
discuss some technical crucial details like thermalization and volume 
scaling. I give a general perspective by discussing Umbrella Methods 
and the Multicanonical approach.
\end{abstract}

\vfill
Lectures given at the $1996$ Budapest Summer School on Monte Carlo 
Methods.
\vfill
\begin{flushright}
	cond-mat/9612010
\end{flushright}
\newpage

\section{Introduction\protect\label{S-INTROD}}

In the following I will give an introduction to optimized Monte Carlo 
methods and data analysis approaches. We will see that the two issues 
are very interconnected, and we will try to understand why.
I will try to keep the same style one has while lecturing, trying 
really to explain all important points in some detail. Even the 
figures will be mostly copied from my transparencies to this text: I 
hope that at least partially that will fulfill the goal I have in mind. 

\section{Reweighting Methods \protect\label{S-REWMET}}

Reweighting methods are based on simple, basic idea: when you run a 
numerical simulation at a given value of the inverse temperature 
$\beta$ and you measure some set of observable quantities 
$O^{(\alpha)}$ (including the internal energy of the system) you are 
learning far more than simply the value of

\be
\langle O^{(\alpha)} \rangle\ ,
\ee
and your ignorance about it (the statistical error). Expectation 
values of the observable quantities at $\beta$ turn out to be only a 
small part of the information you are gathering (if you store the 
right numbers!).

The partition function of our statistical system at $\beta$ can be 
expressed as the integral over the energetic levels allowed to the 
system

\be
Z_\beta = \int\  dE\ N(E)\ e^{-\beta E}\ ,
\ee
where $N(E)$ is the energy density. {\em An accurate numerical simulation 
at $\beta$ gives you information about $N(E)$,} and this information 
can be used in many ways we will discuss in the following. By 
accurate numerical simulation we mean here that in order to make the 
information about $N(E)$ meaningful we need a large sample, and that 
the problem of controlling the statistical and systematic errors is 
here non-trivial. This is also the path to the definition of improved 
Monte Carlo methods like {\em tempering}, that we will discuss in the 
following.

In the following we will be discussing reweighting techniques (also as 
an introduction, as we said, to improved Monte Carlo method, to make 
clear the logical path that leads us there).  After the work of 
\cite{FETAL}, \cite{MARINA} we will discuss first about the simple 
Ising model and the Lee-Yang theorem in (\ref{SS-LEEYAN}) and then 
about the existence of a phase transition in a $4$ dimensional $SU(2)$ 
Lattice Gauge Theory in (\ref{SS-COMZER}).  By doing that we will give 
a very schematic definition of a Lattice Gauge Theory.  We will next 
discuss the use of this approach for improving the quality of the 
analysis of numerical data.  We will introduce hystogramming in 
(\ref{SS-DATANA}) (this is a classical development, based on classical 
work in molecular dynamical simulations by \cite{SAL}, 
\cite{CHE}, \cite{MCD}, and on the mo\-re recent work contained in 
\cite{FETAL}, \cite{MARINA} and \cite{FERSWA}), and the work on 
multi-hystogramming of \cite{FERSWB} in (\ref{SS-MULHYS}).

So, in order to summarize again the physics side of the next section, 
we will start by discussing Lee-Yang zeroes in the $3d$ Ising model, 
by clarifying a few crucial issues about phase transitions.  We will 
discuss how to compute critical exponents from there.  Then we will 
discuss about the $SU(2)$ Lattice Gauge Theory zero analysis, and we 
will see how that helps in establishing that {\em confinement of 
quarks in colorless particles survives in the continuum limit}.  At 
last we will give details about data analysis.  We will also try to 
clarify the path that will eventually lead this technique to be 
promoted, from a data analysis tool, to a (sometimes very effective) 
simulational method.

\subsection{Lee-Yang Zeroes \protect\label{SS-LEEYAN} }

Someone interested in the numerical study of critical phenomena 
should always consider as fundamental the fact that phase transitions 
only exist in the infinite volume limit. On the finite volume (i.e. 
inside all of our computers) there are no phase transitions. Let us 
start by clarifying this point a bit.

We are working in the complex $\beta$ (or $T$) plane.  We consider a 
compact configuration space (spin variables cannot diverge: the Ising 
case with $\pm 1$ values is a very good example, and also a compact 
$SU(N)$ gauge theory with $SU(N)$ matrices is): what we will discuss 
can be also proved under far more general assumptions, but that makes 
things more evident.  The absolute value of the partition function 
$Z_\beta$ is limited from above by

\be
   \left| Z_\beta \right| \equiv 
   \left| \int \{ dC \} e^{-\beta H(\{ C \})}  \right|
   \le V_C e^{|\beta| |H|}
   \ ,
\ee
where the integral is over the configuration space $dC$, 
$V_c \equiv \int \{ dC \}$ is the volume of the configuration space 
and by $|H|$ we denote the maximum value $H(\{ C \})$ can assume when 
considering all possible configurations:

\be
  \left| H \right| \equiv \max_{\{ C \}}\Bigl(H(\{ C \})\Bigr)\ .
\ee
This is also true for complex $\beta$ values (in the case where the 
spin variables can take only discrete values $Z$ is a linear 
combination of exponentials). 

\begin{figure}
  \epsfxsize=400pt\epsffile[28 30 566 311]{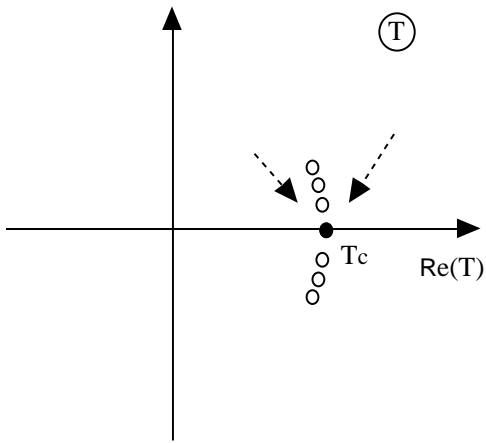}
  \caption[1]{
    The complex $T$ plane. Zeroes at finite volume (empty dots) pinch 
    the real axis at $T_c$ (filled dot) in the infinite volume limit.
    }
  \protect\label{F-FIGLY1}
\end{figure}

What are the properties of the partition function $Z_\beta$?  A 
reasonable $Z_\beta$, which is supposed to describe a physical system, 
is an analytic function in the $\mbox{Re}(\beta)>0$ half of the 
complex plane.  And what happens to the free energy, $-\frac{1}{\beta 
V}\log Z_\beta$?  If $Z$ is an analytic function {\em $\log Z_\beta$ 
can be singular only where $Z_\beta=0$}.  That makes clear the role of 
the {\em zeroes of the partition function}.  For $T$ (and the field 
$h$) belonging to $R^+$ $ Z_\beta$ is a sum of positive contributions, 
and cannot have zeroes in the finite volume $V$.  But, as fig.  
(\ref{F-FIGLY1}) should visualize, in the $V\to\infty$ limit zeroes 
that are located for finite $V$ at complex values of $T_0\in C$ can 
approach the real $T$ axis, generating a non-analyticity of the free 
energy.

The same kind of reasoning can be applied to the behavior in the 
magnetic field $h$: here  for the Ising model the {\em Lee-Yang} 
theorem holds: {\em The zeroes of the partition function are located on 
the imaginary $h$-axis, or on the unit circle of the complex activity 
plane.} In the finite volume there is a finite number of zeroes, and 
in the infinite volume limit the zeroes condense on part of the unit 
circle.

There are no theorems about constraining the zeroes in the complex 
$\beta$-plane. The main theoretical results on this issue have been 
obtained in \cite{ITPEZU}:

\begin{itemize}
\item The complex zeroes close to the (to be) critical point 
accumulate on curves;
\item In $2d$ they cross the real axis at $T_c$ at right angle.
\end{itemize}

We will assume that the same situation holds (with a generic crossing 
angle $\pi-\phi$) in $d>2$. We expect two lines of zeroes in the 
upper and lower positive $T$ complex plane (symmetric because of the 
reality properties of the partition function $Z$), that pinch the 
real $T$ axis at $T_c$. The singular part of the free energy above and 
belove the transition has the form

\be
  f_{\mbox{singular}}^{(\pm)} \simeq A^{(\pm)}
  \Bigl( \frac{T-T_c}{T_c} \Bigr)^{2-\alpha}\ ,
\ee
where $\alpha$ is the usual critical exponent and $A^{(+)}$ and 
$A^{(-)}$ are the specific heat critical amplitudes.  Matching the 
lines of zeroes gives the condition

\be
  \tan\Bigl( (2-\alpha)\phi  \Bigr) =
  \frac{\cos(\pi\alpha)-\frac{A^{(-)}}{A^{(+)}}}{\sin(\pi\alpha)}\ .
\ee
In $2d$ $\phi=\pi/2$ gives that $A^{(-)}=A^{(+)}$, as it is.

So, we have scaling laws for the position of complex zeroes of the 
partition function. Finite size dependence can be derived in the 
usual way, and we will be able to try a numerical experiment to 
determine the critical behavior.

The numerical simulation will be based, as we told, on the fact that 
from a Monte Carlo simulation at a fixed $\beta$ value we can gather 
information about other values of $\beta$ (even complex values).  
Running directly simulations at complex $\beta$ values is far from 
straightforward, and we will find this approach quite direct: the same 
approach will lead us to introduce very powerful Monte Carlo methods.  
Here we will use the method to compute zeroes of $Z_\beta^{(V)}$ at 
$\beta_0=(\Real(\beta_0),\Imag(\beta_0))$, with small 
$\Imag(\beta_0)$ (this is the region that is more interesting 
from the scaling point of view and, luckily enough, is also the one 
that we can access by numerical methods). When we will apply the 
method to real, small $\beta$ increments we will get an useful tool 
for data analysis.

We can start being specific (following  \cite{MARINA}), and 
consider the $3d$ Ising model, with spin variables $\sigma_i=\pm 1$, 
$i$ a $3d$ label of a lattice site, and an action $S$

\be
  S = - \sum' (\sigma_i \sigma_j - 1)\ ,
\ee
where the primed sum runs on first neighboring spin couples on a 
simple cubic lattice. The partition function $Z_\beta$ is written as

\be
  Z_\beta = \sum_{\{C\}}e^{ -\beta S( \{C\} ) } \ .
\ee
At the time of this work an exact solution had been obtained for 
lattices of size up to $4^3$, \cite{PEARSO}: by exact enumeration one 
is counting in this case $O(10^{19})$ configurations! Already a 
lattice of $5^3$ sites cannot be exactly enumerated with today 
computers. As we said our statistical technique will be based on the 
fact that we can express the partition function as a sum over the 
energy levels of the system:

\be
  Z_\beta = \sum_E N(E) e^{-\beta E}\ ,
\ee
where in our case $E=0$, $\ldots$, $d^3$. The instructions are: run 
your Monte Carlo simulation at $\tilde{\beta}$, and record the 
normalized energy distribution function $F_E(\tilde{\beta})$ (this is 
the number of configurations you find at each energy value, 
normalized to one). One has that

\begin{eqnarray}
     \label{E-NF}
	 & F_E(\tilde{\beta}) & 
	 = \frac{f_E(\tilde{\beta})}{Z(\tilde{\beta})} 
	 = \frac{N_E\  e^{\tilde{\beta}E}}{Z(\tilde{\beta})}\ ; \nonumber \\
	\sum_E & F_E(\tilde{\beta})&  = 1\ .
\end{eqnarray}
Now if we compare two different $\beta$ values (we have run at $\tgb$ 
and we are interested in results at $\beta$) we see from 
(\ref{E-NF}) that

\be
  \label{E-DUE}
  N_E = F_E(\beta) Z(\beta) e^{\beta E} 
  =  F_E(\tgb) Z(\tgb) e^{\tgb E}\ .
\ee
We can notice already now that if we stop here, assume that we are 
dealing with two real $\beta$ values, and we use our best numerical 
estimate for the partition function, we get the reweighting formula

\be
  \label{E-FIRST}
  F_E(\beta) = F_E(\tgb)\ 
  \frac{e^{(\tgb-\beta)E}}{\sum_E F_E(\tgb)e^{(\tgb-\beta)E}}\ :
\ee
as we will see better in the following, and we are only anticipating 
here, from the simulation at $\tgb$ we can get expectation values at 
$\beta$ (if $\beta$ is close enough to $\tgb$, and the statistical 
accuracy is good enough not to make the exponential suppression kill 
the signal). But at the moment let us go back to the complex zeroes, 
and rewrite (\ref{E-DUE}) as

\be
  \label{E-FF}
  F_E(\beta) = F_E(\tgb)\ e^{(\tgb-\beta)E}
  \frac{Z(\tgb)}{Z(\beta)}\ .
\ee
Summing over energies and using the normalization condition of 
(\ref{E-NF}) we find

\be
  \label{E-ZZ}
  \frac{Z(\beta)}{Z(\tgb)} = \sum_E F_E(\tgb)\ e^{(\tgb-\beta)E}\ .
\ee
So, we are running a Monte Carlo simulation at $\tgb$, and we are 
computing by a numerical estimate $F_E(\tgb)$.  We want to obtain 
information about the analytic structure of $Z_\beta$ at $\beta \equiv 
\eta + i \xi$.  The exponential factor in (\ref{E-ZZ}) will give, for 
complex $\beta$, two contributions: the first will be an oscillating 
factor, that is non-zero for $\Imag(\beta)\ne 0$,

\be
  \cos(E\xi) + i \sin(E\xi)
\ee
(and we remind that $E$ is here a number of order volume, not of order 
one). The other contribution is the exponential damping

\be
  e^{-(\eta-\tgb)E}\ :
\ee
since $E=O(V)$ the damping is severe. Since we are looking for zeroes 
and we know we cannot get zeroes on the real axis, it is a good idea 
to compute

\bea
  & &
  \frac{Z_\beta}{Z_{\sReal\beta}} =
  \frac{Z_\beta}{Z_{\tgb}} 
  \left\{  \frac{Z_{\sReal\beta}}{Z_{\tgb}}  \right\}^{-1} \label{E-BETTER}
  \\ \nonumber
  &=&
  \frac
  {\sum_E F_E(\tgb)e^{-(\eta-\tgb)E}
  \left(\cos(E\xi) + i \sin(E\xi)\right)}
  {\sum_E F_E(\tgb)e^{-(\eta-\tgb)E}}\ .
\eea
An easy way for a numerical search of zeroes of (\ref{E-BETTER}) is 
to look for minima of

\bea
  & &\left| \frac{Z_\beta}{Z_{\sReal\beta}} \right|^2 =\\
  \nonumber
  & &\frac
  { \left( \sum_E F_E(\tgb)e^{-(\eta-\tgb)E}\cos(E\xi) \right)^2 
    + \left( \sum_E F_E(\tgb)e^{-(\eta-\tgb)E}\sin(E\xi) \right)^2 }
  { \left( \sum_E F_E(\tgb)e^{-(\eta-\tgb)E} \right)^2}\ .
\eea
The numerical simulations of \cite{MARINA} were using 
volumes from $N^3=4^3$ (in order to check the exact solution) to $8^3$. 
$\tgb$ was chosen as close as possible to the actual location of the 
zero in order to minimize the exponential damping. 

We have already said that $\cos(E\xi)=\cos(eV\xi)$, where $e$ is the 
energy density, of order $1$, is highly oscillating, and makes 
impossible to compute the location of zeroes with large imaginary 
part. One nice fact is that by finite size scaling we expect that the 
distance $\delta_N$ of the first zero on a lattice of linear size $N$ 
scales as

\be
  \label{E-HHH}
  \delta_N \simeq N^{-\frac{1}{\nu}}\ :
\ee
Equation (\ref{E-HHH}) can be used to estimate $\nu$ from the rate at 
which zeroes approach the real axis.  It also tells us that the 
oscillations due to the cosine term do not increase like $N_d$, but 
better like $L^{d-\frac{1}{\nu}}$: an exponent of $1.4$ instead of $3$ 
for the $3d$ Ising model.  This helps.

\begin{figure}
  \epsfxsize=400pt\epsffile[28 30 566 320]{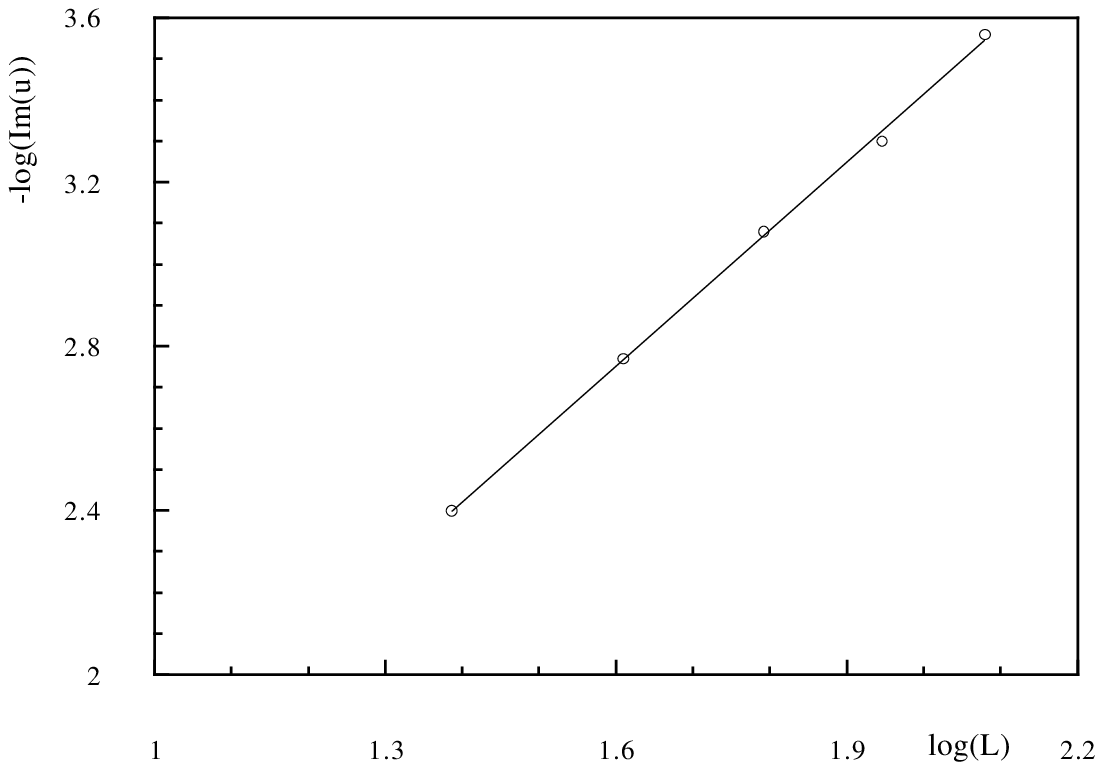}
  \caption[1]{
    $-\log\Imag u^0_N$ versus $\log L$.
    }
  \protect\label{F-FIGLY2}
\end{figure}

In fig. (\ref{F-FIGLY2}) we show the scaling of the distance of the 
first zero from the real axis (one can do the same for farther zeroes, 
but the precision is smaller). We use here the variable $u\equiv 
e^{-4\beta}$. We denote by $u^i_N$ the position of $i$-th zero on a 
lattice of linear size $N$, and plot $u_N^0$ versus $N$ in double log 
scale (the figure here is not precise, and it is only meant as a 
graphical reconstruction of the data: the reader interested in raw 
numbers should consult directly \cite{MARINA}.

The remarkable linearity of the plot denounces a good scaling behavior 
already for small lattices. From these data (numerical archeology at 
today, but we are discussing about the method, not about the numbers!) 
one gets the reasonable estimate $\nu\simeq .620\pm .010$ (the best 
estimate for $\nu$ at the time of this simulation was the far superior
$\nu\simeq .631\pm .001$ by Le Guillou and Zinn-Justin). But with this 
method, for example, we can get a very straightforward measurement of 
the ratio of the critical amplitudes (a quantity that is not easy to 
obtain in the series expansion approach). We measure, as we have 
discussed before, the angle $\phi$ at which the complex zeroes depart 
from the real $u$ axis in the infinite volume limit. One has that

\be 
  \tan\left((2-\alpha)\phi\right) = 
  \frac{\cos(\pi\alpha)-\frac{A^{(+)}}{A^{(-)}}}{\sin(\pi\alpha)}\ .  
\ee 
So, for example, in \cite{MARINA} we got an angle of $55.3\pm1.5$ 
degrees, and an amplitude ratio of $0.45 \pm 0.07$.  This is an 
accurate and reasonable result.  The method works\footnote{Obviously 
there are more recent simulations that follow these lines and are more 
precise: see for example \cite{BHANOT}, and references therein.}!

\subsection{Complex Zeroes in a Non-Abelian Four Dimensional Lattice 
Gauge Theory \protect\label{SS-COMZER} }

In the former section we have described the method one can use to 
locate complex zeroes of the partition function. We will discuss now 
the physical problem for which this technique was first introduced by 
(\cite{FETAL}): I am basically taking this chance to give a ten 
lines crash course in Lattice Gauge Theories, LGT (that here will mainly be 
a fancy Statistical Mechanics, constructed by exploiting a powerful 
symmetry). We will be discussing about locating complex zeroes in a 
non-abelian $4d$ LGT, i.e. about one of the ways of getting numerical 
evidence to show that there is no deconfining phase transition in the 
infinite volume limit. While writing these notes I do not know what 
will Peter Hasenfratz decide to include in his contribution to this 
volume (\cite{HASENF}): in any case it will be an interesting reading 
for the physicist interested to LGT. If in need of more material, you 
could look at one of the books that are available, \cite{ROTHE,MONMUN}, 
to the classical lecture notes by \cite{KOGUT}, to the Les 
Houches notes by \cite{PARLGT}. 

\begin{figure}
  \epsfxsize=400pt\epsffile[28 30 566 320]{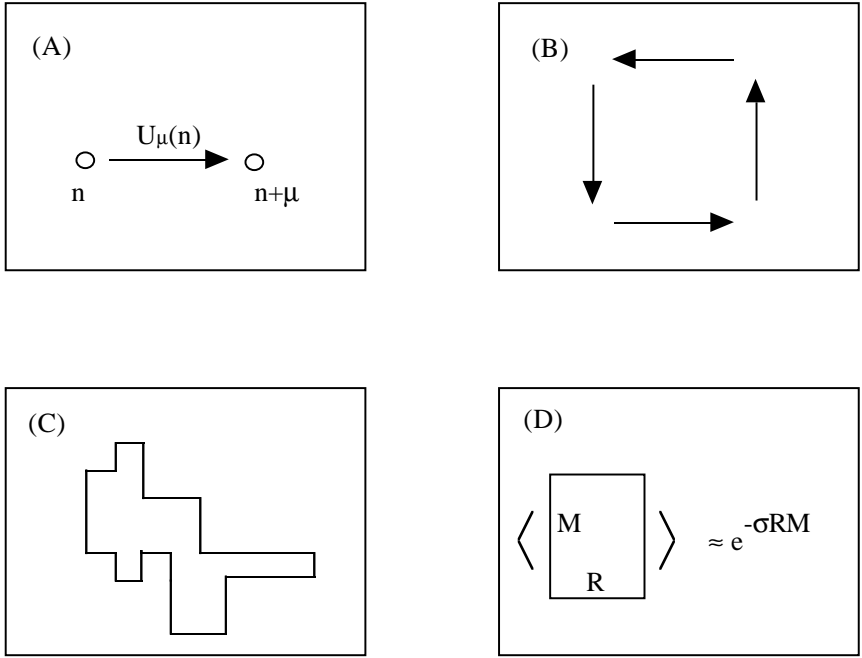}
  \caption[1]{
    From gauge variables to Wilson loops. See text.
    }
  \protect\label{F-FIGGT1}
\end{figure}

In a Lattice Gauge Theory variables leave on {\em links} (as opposite 
to sites for a normal Statistical Mechanics) of a $d$-dimensional 
lattice (simple cubic, for simplicity, in the following), and we will 
label them by $U_\mu(n)$, where $n$ is a $d$-dimensional site label, 
and $\mu$ denotes one of the lattice directions (we show the link 
variable in fig. (\ref{F-FIGGT1}.A)). Different gauge theories are 
characterized by different kind of variables. In the relevant case of 
Quantum Chromo Dynamics, the theory of strong interaction of 
elementary particles, they are $SU(3)$ matrices (here we will consider 
a simpler theory with many similar features, the one of $SU(2)$ 
$2\cdot 2$ matrices). In the case of a $SU(M)$ gauge group $U$ is a 
$M\cdot M$ matrix with $U\cdot U^\dagger=1$, and $det(U)=1$.

The Boltzmann equilibrium probability distribution can be written as

\be
  \label{E-BOL}
  P_B \simeq e^{\beta\sum_P U_P}\ ,
\ee
where the sum runs over all {\em plaquettes} (the smallest closed 
circuits, see (\ref{F-FIGGT1}.B)) of the lattice, and 

\be
  U_P \equiv U_\mu(n) U_\nu(n+\hat{\mu}) 
  U^\dagger_\mu(n+\hat{\nu}) U^\dagger_\nu(n)\ ,
\ee
i.e.  one takes on each elementary plaquette the product of ordered 
link matrices (by defining $U_\mu(n)=U^\dagger_{-\mu}(n+\hat{\mu})$).

This theory has a dramatically large invariance, known as {\em gauge 
invariance}: if we pick up arbitrarily a group element $g(n)$, that we 
can choose independently on each site, and we transform all the link 
variables under

\be
  U_\mu(n) \to g(n) U_\mu(n) g^\dagger(n+\hat{\mu})\ ,
\ee
the action of the theory does not change (and do not change all 
observable quantities, i.e.  products of links on closed loops, see 
(\ref{F-FIGGT1}.C)). In the Ising model the crucial $Z_2$ symmetry is 
only a {\em global} symmetry: the theory is symmetric under inversion 
of all $\sigma(n)$ variables. Here, on the contrary, we have the 
right to select an independent frame rotation at each lattice point, 
and the theory does not change. Such gauge symmetry is exact in the 
lattice theory. This is the crucial step of the Wilson approach to 
LGT: in the same way in which the $2d$ Ising model Onsager solution is 
described by Majorana fermions at the critical point, where the 
correlation length becomes infinite and details of the lattice 
structure are forgotten, the critical limit of lattice QCD is the 
usual continuum QCD, the theory of interacting quarks and gluons. The 
fact that continuum gauge invariant is exactly preserved on the 
lattice (as opposed, for example, to Lorentz invariance, that is 
obviously broken by lattice discretization) is a crucial point of the 
approach.

We have said already, and we will not enter in 
details, that products of link variables on closed loops are the 
observable quantities.

We also remark that the inverse temperature $\beta$ that appears in 
the Boltzmann distribution (\ref{E-BOL}) is connected to the coupling 
constant of the continuum gauge theory one recovers in the limit of 
small lattice spacing:

\be
  \beta_{StM}\simeq g_{GT}^{-2}\ ,\ \ T_{StM}\simeq g_{GT}^{2}\ .
\ee
The theory has a phase transition at $T=0$ (here $g\to 0$), where the 
correlation length diverges (exponentially in $\frac{1}{T^{2}}$). As 
usual in this continuum limit the lattice structure becomes 
irrelevant. 

For high values of $T$ it is easy to prove the relation we have 
depicted in (\ref{F-FIGGT1}.D): a Wilson loop (the product of oriented 
link variables over a closed loop) of large size $R\cdot M$ decays as 
the exponential of minus the {\em string tension} $\sigma$ times the 
area of the loop in the confined region.  If quarks are confined and 
cannot be separated out from color singlet states (the physical 
particles, mesons and baryons) we have an area decay of large Wilson 
loops. So, the nice part is that, as we said, it is easy to prove that 
Lattice QCD is confined in the high $T$ region (where the theory is 
very different from the continuum theory). The bad  part is that one 
can prove that for all LGT, including the Lattice QED: since 
electrons are known to be free in the continuum theory, if everything 
goes right in this case something will have to happen. What happens 
for example in the case of QED is that a finite $T$ phase transition 
separates two regions, the confined, unphysical one, and the 
deconfined physical theory. One has to show that the same does not 
happen in Lattice QCD, and that the theory does not undergo a phase 
transition that would destroy confinement.

\begin{figure}
  \epsfxsize=400pt\epsffile[28 30 566 320]{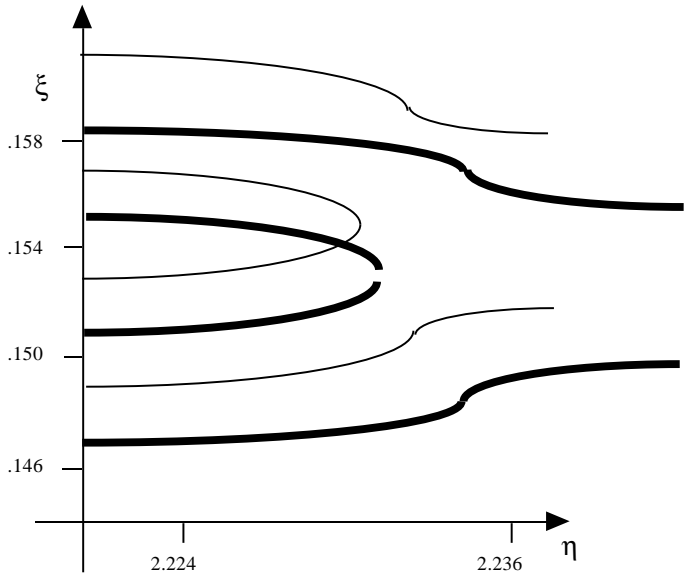}
  \caption[1]{
    Lines of zeroes of the imaginary part (thin curves) and of the real 
    part (thick curves) of the ratio of partition functions.
    }
  \protect\label{F-FIGGT2}
\end{figure}

Our analysis of lattice zeroes was studying this problem. The 
technique is exactly the same we have described in the former section. 
In fig. (\ref{F-FIGGT2}) we draw curves with

\be
  \Real\left(  \frac{Z(\eta+i\xi}{Z(\eta)}   \right)=0\ ,
\ee
and curves with

\be
  \Imag\left(  \frac{Z(\eta+i\xi}{Z(\eta)}   \right)=0
\ee
(for the exact curves the reader should consult the original 
reference, \cite{FETAL}).  It is clear from the figure one can 
determine the zero with good precision.  Since one finds a zero quite 
far from the real $\beta$ axis, and it does not approach the real axis 
for larger lattice sizes, one does not expect a real phase transition, 
but is measuring a transient phenomenon that is irrelevant as far as 
the real continuum limit is concerned.  It is clear that the evidence 
we have discussed here is the same one exploits when using finite size 
scaling techniques, looking for example at the behavior of a peak of 
the specific heat.  It is interesting that one can directly study the 
position of complex zeroes of the partition function.

\subsection{Data Analysis \protect\label{SS-DATANA} }

In equation (\ref{E-FIRST}) we have already seen that our approach 
can be used to deduce information at $\beta$ after running a 
simulation at $\tgb$. We can generalize  (\ref{E-FIRST}) by noticing 
that we can sample also magnetizations $m$ (so we can reconstruct all 
moments of $m$). After each measurement we write down the energy and 
the magnetization of the configuration (that we assume to be at 
thermal equilibrium). We have that

\be
  \label{E-WITHH}
  F_{E,m}(\beta,h) = F_{E,m}(\tgb,\th)\ 
  \frac{e^{(\tgb-\beta)E+(\th-h)m}}{\sum_{\tE,\tm} 
  F_{\tE,\tm}(\tgb,\th)e^{(\tgb-\beta)E+(\th-h)m}}\ .
\ee
In \cite{FERSWA} one finds a very nice picture showing how well the method 
can work for example for the case of the $2d$ Ising model.

\begin{figure}
  \epsfxsize=400pt\epsffile[28 30 566 320]{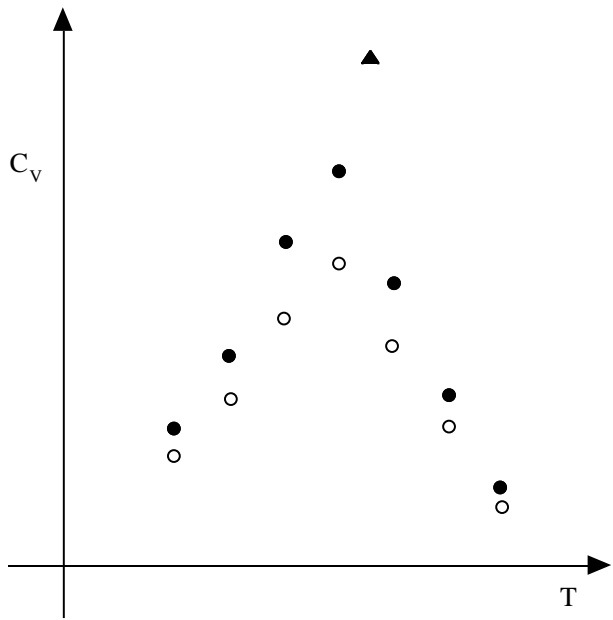}
  \caption[1]{
    The specific heat versus $T$ for a typical finite size scaling 
    study suffering of troubles that can be cured by reweighting. 
    Empty dots are for the smaller lattice, filled dots for the 
    larger lattice, and the triangle for the point we did not measure 
    and we should have measured.
    }
  \protect\label{F-FIGDA1}
\end{figure}

The method we have described here is very useful, for example, when 
one wants to measure the finite size scaling behavior of physical 
observable quantities.  Let us consider for example the specific heat 
$C_V$.  The maximum value $C_V$ takes on a finite lattice of linear 
size $L$, $C_V^{\max}(L)$ scales at a critical point as

\be
  C_V^{\max}(L) \simeq L^{\frac{\alpha}{\nu}}\ .
\ee
The main problem is in the fact that we only measure for a discrete 
set of values of the temperature $T$ (by normal MC or by using a 
$T$-changing Monte Carlo procedure, see later in this notes). A 
priori we do not know at which value of $T$ on a given lattice the 
specific heat takes its maximum value, and such $T$ value depends on 
$L$:

\be
  T^{\max}(L)\  |\    C_V( T^{\max}(L),L) \equiv C_V^{\max}(L) \ 
  \mbox{depends on}\ L\ .
\ee
It can be difficult to find the correct value of $T^{\max}(L)$: it is 
a delicate fine tuning process that has to be repeated for each 
different $L$ value. A wrong determination of $T^{\max}(L)$ can 
generate a very misleading effect. Let us look at figure 
(\ref{F-FIGDA1}). The empty dots represent ({\em gedanken}) 
measurements of the specific heat taken with a linear size $L$, at 
the set of $T$ values which appear in the figure. The filled dots are 
measurements on a lattice of larger linear size $L'$, taken at the 
same $T$ values. One would assume that the finite size scaling 
behavior is given by the scaling of the two measured points with 
higher value of $C_V$. But maybe the real maximum, on the larger 
lattice, is where the triangle is: there the scaling could be very 
different from the one we found on the points we measured, but sadly 
we did not measure on this point. We want to stress that this effect 
creates real practical problems in numerical simulations.

The pattern of data analysis we have discussed here solves this 
important problem.  Statistical error can be kept under control (by 
using for example jack-knife and binning techniques), and 
numerical studies of scaling become (without further computational 
expenses, since we already had the information) a real quantitative 
tool.

\subsection{Multi-Hystogramming \protect\label{SS-MULHYS} }

The idea of reweighting can be pushed forward. \cite{FERSWB} 
introduced the {\em multi-hystogramming}. The crucial step is to 
realize that you can patch data from different simulations at 
different $\beta_k$ values to reconstruct all of the relevant $\beta$ 
region. 

So, we have to sum up histograms.  The delicate point is how to sum 
them up, i.e.  how to determine the weights to use in constructing 
linear combinations of the different entries.  The way discussed in 
\cite{FERSWB} is to determine the weights by minimizing the 
statistical uncertainty over the final estimate for $P_\beta(E)$.  Let 
us call $N_k(E)$ the data histogram for the $k$-th Monte Carlo run, at 
$\beta_k$.  Let us define $\theta_k \equiv 1+2\tau_k$, where $\tau_k$ 
is the estimated correlation time at $\beta_k$, and $n_k$ the number 
of sweeps used in the $k$-th Monte Carlo run.  One finds that the 
parameters $\{g_n\}$ can be determined self-consistently by iteration 
from

\bea
  P_\beta(E) &=& \frac
  { \sum_{k=1}^R \ N_k(E) \ e^{\beta_k E}\         \theta_k^{-1} }
  { \sum_{j=1}^R \ n_j    \ e^{\beta_k E - g_j}\   \theta_k^{-1} } \ ;
\nonumber \\
  e^{g_j} &=& \sum_E  P_\beta(E) \ .
\eea
We have denoted by $R$ the number of Monte Carlo runs. The method 
works well, and at today it can be considered as a standard analysis 
tool.

\section{Improved Monte Carlo Methods\protect\label{S-TEMPER}}

We will discuss here about improving Monte Carlo methods.  We will be 
mainly talking about {\em tempering}, by \cite{TEMPER} (see section 
(\ref{SS-TEMPER})), where you add a second Markov chain to the usual 
Metropolis chain: you let $\beta$ vary, trying in this way to make 
easier for the system to move across deep free energy valleys 
separated by high free energy barriers.  We will also try to discuss 
about general issues like {\em thermalization}, that are of crucial 
importance already when discussing simple Monte Carlo methods, and 
that turn out to be even more delicate issues here.

Eventually one is looking for a very effective, very simple simulation 
method.  Somehow when you start the numerical study of a statistical 
system you work at two different levels.  At first you run a quick and 
not very clean Monte Carlo, to understand the first physics 
ideas\footnote{The phrase typically used by G. Parisi to describe this 
approach is {\em Il buon giorno si vede dal mattino}, that I would 
translate in english with {\em It is already early in the morning that 
you can recognize a good day from a bad one.}}.  Only after that 
you set up complex simulational procedures, data analysis, error 
determination.  It would be nice if the first phase we have just 
describe could already be based on something more powerful that the 
usual Metropolis approach: I hope the reader will be convinced that 
maybe the {\em Parallel Tempering} approach (see section 
(\ref{SS-PARTEM})) is the good candidate for this role.  In Parallel 
Tempering there are no parameters to be tuned, no difficult choices to 
be done (but for the selection of a set of $T$ values that can be done 
with some large freedom): it looks like the good thing.

\subsection{Simulated Tempering\protect\label{SS-TEMPER}}

We will introduce here {\em Simulated Tempering}, \cite{TEMPER}, an 
improved Monte Carlo method that turns out for example to be very 
effective for simulating spin glasses (for further studies and 
applications of tempering see \cite{FE,CO,VI,KE,CA}).  Later on we 
will discuss about {\em Parallel Tempering}, 
\cite{TESI,GEYER,HUKU1,HUKU2}, that turns out to be a better and 
simpler method (having at the same time these two advantages is a 
quite rare and appreciated feature).  So we will discuss now some 
complex matter that we will eventually not need in the practical 
implementation of the method: we will do that since it helps in 
understanding some of the physical mechanisms that govern the scheme.  
These mechanisms are common to the most promising parallel tempering 
scheme.  Parameter tuning is not needed in parallel tempering.

Simulated tempering is a global optimization method: it can be seen 
as an annealing, generalized to $T\ne 0$, with a built-in scheduling. 
This can be of large practical importance, since setting up the 
schedule is one of the most difficult and time taking parts of an 
annealing simulation. Tempering is very similar to annealing, but 
after the initial thermalization period the field configuration is 
always at Boltzmann equilibrium at one of the allowed $\beta$ values. 
This phrase, that is a bit mysterious at this point, is important, 
and will be clarified in the following.

There are many related techniques, strictly connected methods and 
needed introductive material.  In first one needs to know about 
generalities of Monte Carlo methods (see for example the lectures in 
this book by \cite{KRAUTH}).  Theory of multiple Markov chains is the 
mathematical basis one needs to clarify the theoretical aspects of the 
method (\cite{TESI,GEYER,HUKU1,HUKU2}).

Strictly related to tempering are the {\em scaling} approaches based 
on {\em Umbrella Sampling}, \cite{TVA,TVB,GV,VAL}: we have already discussed 
the issue when illustrating data analysis reconstruction.  It has to 
be noticed that many of the ideas we are applying now to numerical 
Statistical Mechanics and Euclidean Statistical Field Theory have been 
elaborated many years ago in the context of physics of liquids and of 
Molecular Dynamics.

{\em Multicanonical Approaches} by \cite{BE31,BE32,BE1,BE2} 
are also strictly related to tempering, and we will also 
discuss about them in the following.  Multicanonical methods are more 
ambitious and in many situations potentially more powerful than 
tempering (but they are a bit more complex): they can try and deal 
with first order phase transitions, that is not something you want to 
do with tempering (that works well for continuous phase transitions). 
We also note that different kind of tempering-like approaches have 
been proposed, for example in \cite{STATEM}.

As we have discussed from the start of these notes Tempering has been 
built on reconstruction methods, \cite{FETAL,MARINA,FERSWA,FERSWB}: if 
we can use data at $\tgb$ for learning about expectation values at 
$\beta$ maybe we can also use this information for speeding up the 
dynamics itself (again, see \cite{KRAUTH} for an introduction to Monte 
Carlo simulations and Markov chains). When you run a long simulation 
you could know much more than you believe (or much less, but we will 
discuss about that when talking about thermalization and correlation 
times).

Let us start with describing {\em Simulated Tempering}. We want to 
equilibrate our statistical system with respect to the Boltzmann 
distribution

\be
  \label{E-BOLTZMANN}
  P(\{\sigma\}) \simeq e^{-\beta S(\{\sigma\})}\ .
\ee
We choose a {\bf new} probability distribution, with an enlarged 
number of variables:

\be
  \tP(\{\sigma\},\{\Sigma\})\ ,
\ee
such that, for each given choice of the $\{\Sigma\}$, $\tP$ is 
Boltzmann with some given choice of $\beta$. {\em We will allow 
$\beta$ to become a dynamical variable:}

\be
  \{\sigma\}\ \to\ (\{\sigma\},\{\beta_\alpha\})\ ,
\ee
$\alpha=1, \ \ldots,\  A$. We have allowed $A$ fixed values for the 
$\beta_\alpha$ variables.

The mechanism we are introducing is very simple: the new equilibrium 
probability distribution is

\be
  P_{eq}(\{\sigma\},\{\beta_\alpha\}) 
  \simeq e^{-\beta_\alpha H(\{\sigma\})+g_\alpha}\ ,
\ee
where $H$ is the original Hamiltonian of the problem. The $g_\alpha$ 
are constant quantities, whose meaning will be elucidated in the 
following. They have to be fine tuned before running the equilibrium 
sampling, and they only depend on the value of $\alpha$ (only one 
$g_\alpha$ is allowed for each $\beta_\alpha$ value).

The probability of finding a given value of $\alpha$ (i.e. the 
probability for a given $\beta_\alpha$ value to occur during the 
run) is:

\be
  \label{E-GF}
  P_\alpha = 
  z_\alpha e^{g_\alpha}\equiv 
  e^{g_\alpha - \beta_\alpha f_\alpha}\ ,
\ee
where $z_\alpha$ is the partition function at fixed $\beta_\alpha$,

\be
  \int\{d\sigma\}e^{- \beta_\alpha H(\{\sigma\})}\ ,
\ee
and $f_\alpha$ is the free energy of the system with fixed 
$\beta_\alpha$. 
(\ref{E-GF}) shows that the free parameters $g_\alpha$ are connected 
to the free energy of the system. For making the probability of 
different $\beta$ values to occur to be equal (i.e. the system to 
visit with the same frequency all allowed $\beta$ values) we need to 
set

\be
  \label{E-BESTG}
  g_\alpha = \beta_\alpha f_\alpha\ .
\ee
Since we do not know th $f_\alpha$ that will amount to use for the 
$g$'s the best available estimate for $f$. We will discuss in the 
following how to produce a good guess for the $g_\alpha$, that can 
also be improved systematically. Since eventually parallel tempering 
will not need any kind of parameters to be fine tuned we will not go 
in details about this issue (explaining what the $g$ are in simple 
tempering is useful for reaching a better physical understanding of 
both tempering and parallel tempering).

\begin{figure}
  \epsfxsize=400pt\epsffile[28 30 566 270]{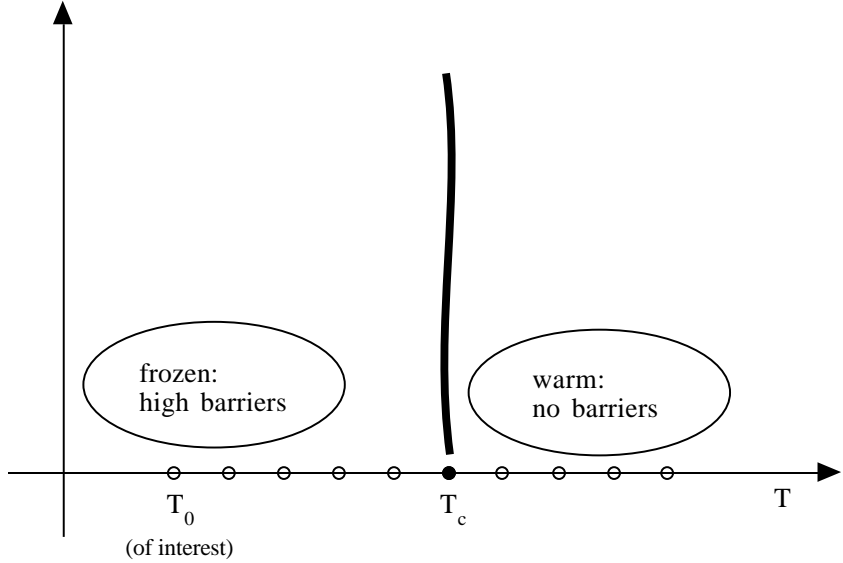}
  \caption[1]{
    The choice of the $T$ values to be used in the tempered 
    simulation of a complex system.
    }
  \protect\label{F-FIGTE1}
\end{figure}

If we are mainly interested in studying the system at $T=\tT$ we will 
allow for a set of $T_\alpha$, for example at constant distance, 
$T_0\equiv \tT$, $T_1\equiv (\tT+\delta)$, $T_2\equiv 
(\tT+2\delta)$, $\ldots$ (see figure (\ref{F-FIGTE1})). We will 
discuss how to select optimal values for $\delta$ (this has to be 
done also in parallel tempering, but one does not need fine tuning). 
After the runs if you want you can use reconstruction schemes to use 
all of the information contained in all samples, but in this kind of 
approach this is typically not necessary (since we are looking for 
thermalized configurations in a complex, cold broken phase, and the 
main information is typically contained in the $T_0$ data).

The main physical ideas are

    \begin{enumerate}
    	\item  The system is frozen at low $T$. High energy barriers 
    	separate different free energy valleys.
    
    	\item  When the system warms up during tempering the free energy 
    	barriers get smaller. They disappear when the system crosses 
    	$T_c$. 
    
    	\item  When it cools down again it will probably explore a 
    	different local minimum.
    \end{enumerate}
    
The method seems to work nicely for second order like phase 
transitions: for that to happen you need the broken state to be 
conformationally similar to the high $T$ one.

We will now analyze in detail a full updating sweep, in order to make 
the procedure clear.

{\bf 1) } Sweep the full lattice, maybe $s$ times ($s>1$ could 
help) and run a full normal Monte Carlo (Metropolis, or cluster, or 
over-relaxed or what you like better) update for the $\{\sigma\}$ 
variables at {\em fixed} $\beta_\alpha$.

{\bf 2) } Propose the update

\be  
  \beta_\alpha \to \beta_{\alpha'}\ ,
\ee
where $\alpha'=\alpha\pm 1$ with probability $\frac12$ (at $\alpha=1$  
and $\alpha=A$ 
we can or not decide not to move with probability $\frac12$, since we 
are not interested in detailed balance in $\beta$ space: see 
\cite{KRAUTH} about the importance of rejected moves).

We accept the update with normal Metropolis weight. Here the factor in 
the exponential only changes because of the change in $\beta$, since 
the configuration energy does not change:

\be
  \label{E-DELTA-S-TEM}
  \Delta S_{\mbox{tempering}} = (\beta_{\alpha'} - \beta_\alpha)
  H(\{\sigma\}) + (g_{\alpha'}-g_\alpha)\ ,
\ee
where both the $H$ and the $(g_{\alpha'}-g_\alpha)$ terms are 
constant. If $\Delta S <=0$ we accept the $\beta$ update, if $\Delta 
S >0$ we accept it if a random number uniformly distributed in 
$(0,1)$ is smaller than $e^{-\Delta S}$.

A good first guess for the $g_\alpha$ can be deduced from 
(\ref{E-DELTA-S-TEM}) (then, as we said, the $g$'s can be 
systematically improved). We can take the $g$ such that in average 
$\Delta S$ is balanced:

\be
  (g_{\alpha'}-g_\alpha) = 
  - (\beta_{\alpha'} - \beta_\alpha)
  \frac
  { \langle H \rangle_{\beta_{\alpha}} + 
    \langle H \rangle_{\beta_{\alpha'}} }
  {2}\ ,
\ee
that holds at first order in $\delta\beta$.  We see that the $g$, by 
balancing for the free energy of the system, do not allow the system 
to follow in the lowest energy state and stay there forever.  The 
first correction is of the form

\be
  C_V^\alpha (\delta\beta)^2\ ,
\ee
and keeping this term of order one implies that in the large volume 
limit $\delta\beta$ has to be kept of order $ C_V^{-\frac12}$, see also 
later.

Now the crucial observation: at each $\beta_{\alpha}$ value (after the 
initial thermalization) the system is always in equilibrium with 
respect to the {\em usual} Boltzmann distribution. The system is {all 
the time} (even at the moment of $\beta$-transitions) at thermal 
equilibrium. 

The analysis of observable quantities is very easy. One just has to 
select all configurations which were flagged by a given $\beta$ value:

\be
  \langle O \rangle_{\beta_{\alpha}} = \frac
  {\sum_{\mbox{All configurations at} \beta_{\alpha}} O_{\mbox{conf}} }
  { \mbox{Number of configurations at} \beta_{\alpha} }\ .
\ee

\begin{figure}
  \epsfxsize=400pt\epsffile[28 30 566 320]{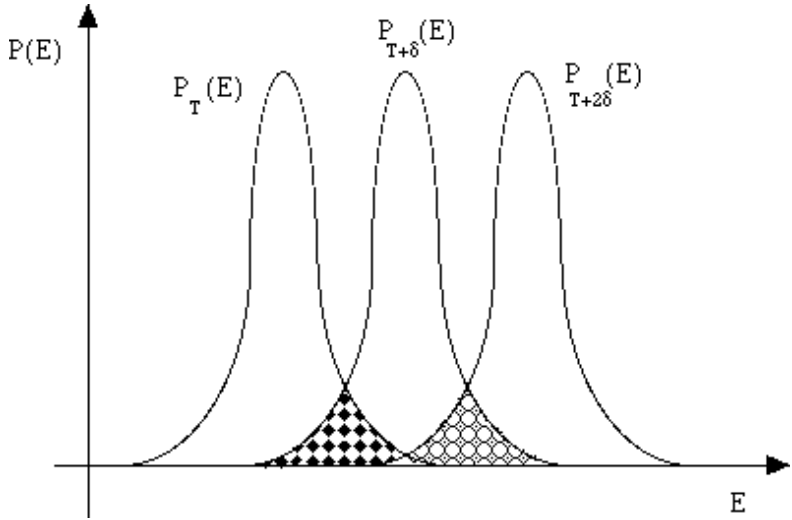}
  \caption[1]{
    $P(E)$ for different values of $T$.
    }
  \protect\label{F-FIGTE2}
\end{figure}

How do we select a good set of $\{\beta_{\alpha}\}$ values? Fig. 
(\ref{F-FIGTE2}) can help to give an idea. Let the first $P(E)$ on 
the left be the one at the lowest $T$ value (the one we are 
interested in thermalizing). The center the one is at $T+\delta$, i.e. 
at the second $T_\alpha$ value. We request that the overlap of the two 
probability distributions is non-negligible. A configuration with an 
energy included in the colored region of the picture is typically a 
good configuration both at $T$ and at $T+\delta$ (it is here that 
problems connected to discontinuities, like in first order phase 
transitions make the method fail). So, $\delta\equiv T_{\alpha+1}- 
T_\alpha$ has to be selected such that {\em there is a non-zero 
overlap among the two energy probability distributions} (as usual in 
Metropolis like methods that is connected to having a reasonable 
acceptance factor, of order $0.5$, for the $\beta_\alpha$ moves).
That should also make more clear why we could like to do more than 
one $\{\sigma\}$ sweeps before updating $\beta$: we want to avoid a 
series of jumps between adjacent $\beta_\alpha$ values, and give
to the system the possibility of moving at fixed $\beta$ at the opposite 
extremum of the $P(E)$ before trying changing $\beta$ again.

We repeat which are the basis of the physical mechanism we are 
interested in.  We are thinking about systems with very high free 
energy barriers.  Warming the system up lowers the barriers, which 
disappear at $T_c$.  When cooling down (always staying at thermal 
equilibrium!)  the system can fall in a completely new valley.

Experimentally the method turns out to work very well for $3d$ 
Edwards-Anderson spin glasses.  It is interesting to note that 
temperature chaos, \cite{KO,KOVE,RI}, should give troubles, but on the 
lattice sizes we have been able to study we did not get them: it is 
interesting to try and understand more about this issue.  In the case 
of $3d$ Edwards-Anderson spin glasses by using parallel tempering (see 
(\ref{SS-PARTEM})) we have been able to thermalize a $16^3$ system down 
to $0.7T_c$ (\cite{MAPARU}).  On a $24^3$ lattice one is able to 
thermalize with a reasonable computer time (we are talking about 
single runs on one disorder sample taking order of one day of 
workstation) down to $0.9T_c$, while for lower $T$ the situation gets 
more difficult to control.

The method does not work, \cite{IOMAPB}, for generic heteropolymers 
(with order $30$ sites, Lennard-Jones potential with quenched random 
couplings \cite{IOMAPA}), and this is probably connected to the fact 
that the glassy transition is in this case of a discontinuous nature: 
for true (small) proteins, on the contrary, the method seems to be 
helpful \cite{HANSMA}.  Also the method has been 
applied among others (\cite{FE,CA}) by \cite{CO} to the $4d$ 
disordered Heisemberg model, by \cite{KE} to the $2d$ EA spin glass 
(by noticing one of the important advantages of the method, i.e.  the 
trivial vectorization and parallelization of the scheme), and by 
\cite{VI} to $CP^{N-1}$ models.

\subsection{Random Field Ising Model\protect\label{SS-RFIM}}

The case study we have done in our first tempering paper, 
\cite{TEMPER}, 
was discussing the {\em Random Field Ising Model}, and it is ideal to 
illustrate in some more detail the main feature of the method: we 
will discuss it here. The model is defined by the Hamiltonian

\be
  H = -\sum_{\langle i,j \rangle}\sigma_i \sigma_j +\sum_i h_i 
  \sigma_i\ ,
\ee
where $\sigma_i=\pm 1$, the sum with two indices runs over first 
neighboring sites on a simple cubic lattice in $d$ dimensions, 
and the local external fields $h_i$ are 
quenched random variables, which take the value $\pm |h|$ with equal 
probability. 

\begin{figure}
  \epsfxsize=400pt\epsffile[28 30 566 150]{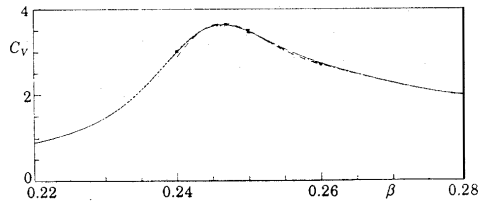}
  \caption[1]{
    The measured specific heat $C_V$ as a function of $\beta$. 
    $\beta=0.26$, in the cold region, is the point on which we focus.
    }
  \protect\label{F-FIGRF1}
\end{figure}

We have taken $d=3$, $V=10^3$ sites and $|h|=1$, sitting close to the 
critical region, on the cold side (we will focus here on studying 
$\beta=.26$).  Fig.  (\ref{F-FIGRF1}) gives an idea about the critical 
region.  The specific heat has a maximum close to $\beta=.25$.  In 
these first runs in most cases we allowed the system to visit only $3$ 
$\beta$ values, ($\beta_\alpha$ $=$ $(.24,.25,.26)$), and sometimes 
$5$ $\beta$ values: for the Edwards Anderson model in the most recent 
runs of \cite{MAPARU} on a $24^{3}$ lattice we use up to $50$ $\beta$
values. The $3$ $\beta$ values we have given above are selected such 
to span from the cold phase to the warm phase. This is the general 
principle: the system has to be allowed to travel, in $\beta$ space, 
from the cold phase, that has a physical interest, to the warm phase, 
where free energy barriers disappear and correlation times are short.

\begin{figure}
  \epsfxsize=400pt\epsffile[28 30 566 200]{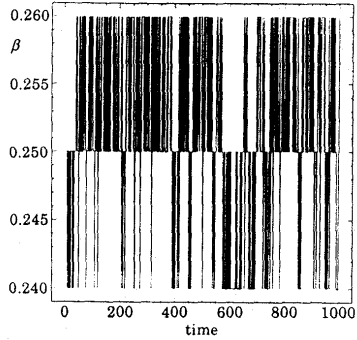}
  \caption[1]{
    $\beta_\alpha$ as a function of the Monte Carlo tempering time.
    }
  \protect\label{F-FIGRF2}
\end{figure}

In fig. (\ref{F-FIGRF2}) we plot the $\beta_\alpha$ values the system 
selects in the course of the dynamics. Notice that the system is not 
getting trapped in the cold or in the warm phase: acceptance of the 
$\beta$ swap is easy, and the system easily travels among the two 
phases.

\begin{figure}
  \epsfxsize=400pt\epsffile[28 30 566 200]{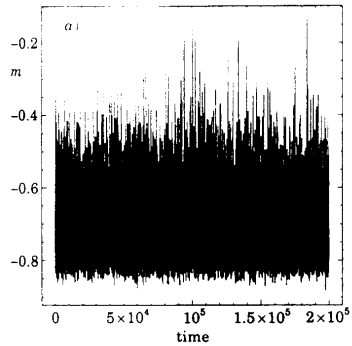}
  \caption[1]{
    Magnetization as a function of Monte Carlo time, at $\beta=0.26$, 
    for normal Metropolis algorithm. The system never tunnels to a 
    positive magnetization value.
    }
  \protect\label{F-FIGRF3}
\end{figure}

\begin{figure}
  \epsfxsize=400pt\epsffile[28 30 566 200]{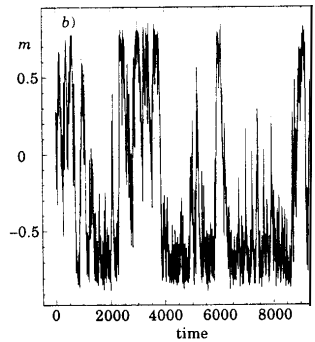}
  \caption[1]{
    Magnetization versus Monte Carlo time for tempering. Here all 
    $\beta$ values are included: the measured magnetization are 
    related to configurations that are at Boltzmann thermal 
    equilibrium with different temperature values.
    }
  \protect\label{F-FIGRF4}
\end{figure}

The magnetization measured from a typical normal, fixed $\beta$ Monte 
Carlo metropolis runs is shown in fig.  (\ref{F-FIGRF3}): here the 
correlation time, since one never sees a flip to the opposite 
magnetization time, seems short.  That taking fig.  (\ref{F-FIGRF3}) 
at face values would be misleading is clear from the magnetizations 
from a tempering run shown in fig.  (\ref{F-FIGRF4}): here tempering 
is allowing the system to flip among the $\pm m$ states, and it is 
clear that configurations with positive $m$ have a relevant weight in 
the equilibrium distribution.

\begin{figure}
  \epsfxsize=400pt\epsffile[28 30 566 200]{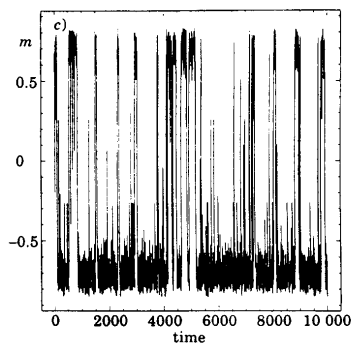}
  \caption[1]{
    As in fig. (\protect\ref{F-FIGRF4}), but we have selected only 
    configurations with $\beta=0.26$.
    }
  \protect\label{F-FIGRF5}
\end{figure}

In fig.  (\ref{F-FIGRF5}) we plot the magnetization of the 
configurations that got a $\beta$  value of $0.26$. Here one 
clearly sees that the state with positive $m$ is allowed at 
$\beta=0.26$ with a probability smaller than for the negative $m$ 
state but definitely non-zero. These last figures give the main idea. 
In a non-disordered Ising model flipping from the minus state to the 
plus state would be irrelevant, if we are not interested in studying 
details of the tunneling dynamics: we know by symmetry that to the 
minus state corresponds an identical plus state. But this is not true 
for a disordered model, where on the contrary the tunnel among 
degenerate ground states that are not related by a trivial symmetry is 
the most interesting part of the dynamics. Here the random quenched 
magnetic field breaks the symmetry, and we want to explore the different 
ground state. Tempering allows us to do that.

This was a very easy numerical experiments, but the steps we have 
described characterize also a more complex settings: for example we 
already told that the numerical simulations we are running now 
(\cite{MAPARU}) on large lattices for the $3d$ Edwards Anderson spin 
glass are quite complex (but they work very well).

Tempering is related to annealing. A trivial extension of annealing to 
non-zero values of $T$ is impossible: annealing gives only 
information about energy, while for $T>0$ we want to deal with the 
free energy. We do not get from annealing any information about the 
entropic structure of the phase space. Tempering can be seen as such a 
generalization.

Tempering could also turn out to be an effective global optimization 
scheme, even if this issue has non been looking in much detail until 
now. Very preliminary unpublished studies are by \cite{ROSE,SHORE}. 
The most important point is that tempering contains a built-in, 
self-implemented schedule. Setting up the schedule is the most 
serious problem with annealing, and tempering does it for us.

\subsection{Parallel Tempering\protect\label{SS-PARTEM}}

Parallel Tempering has been discussed by 
\cite{TESI,GEYER,HUKU1,HUKU2}. We will describe it here. As we already 
said it is so simple that makes many of the details we have discussed 
with tempering only informational (we look at that as a plus). Also, 
the parallel tempering performs dramatically well, for example, on 
finite dimensional Edwards Anderson models.

In figures (\ref{F-FIGTE1}) and (\ref{F-FIGTE2}) we have shown how we 
select the $T$ values we use during the simulation.  Let us say that 
by using the criteria we have defined before we need for example 
$N_{(\beta)}$ values of $\beta_\alpha$.  Now in the parallel tempering 
approach, you run $N_{(\beta)}$ simulations in parallel ( 
$N_{(\beta)}$ different configurations $C_\alpha$ of the system that 
evolve in the same quenched disordered landscape).  Each copy starts 
with assigned a different $\beta$ values.  For example start with

\be
  \beta(C_0)=\beta_0;\ \beta(C_1)=\beta_1;\ \ldots \ 
  \beta(C_\alpha)=\beta_\alpha;\ \ldots \ 
  \beta(C_{N_{(\beta)}}) = \beta_{N_{(\beta)}}\ .
\ee
Now the composite Monte Carlo method is based on two series of steps:

\begin{enumerate}

	\item  Update the $N_{(\beta)}$ copies of the system with an usual 
	Metropolis sweep of 
	\be
	\left\{\sigma\right\}_{C_\alpha}  \ldots
	\left\{\sigma\right\}_{C_{N_{(\beta)}}}\ .
	\ee

	\item  Swap the $\beta$ values.
	
	\begin{itemize}
	
		\item Propose the first $\beta$-swap: the configuration that is 
		at $\beta_0$ can go to $\beta_1$ and that one that is at $\beta_1$ 
		can go at $\beta_0$. Each spin configuration carries a 
		$\beta$-label. Configurations carrying adjacent $\beta$-labels try 
		to swap their labels. You use Metropolis for swapping $\beta$ with 
		the correct probability (see later).
		
		\item Propose the second $\beta$-swap: the configuration that is 
		at $\beta_1$ can go to $\beta_2$ and that one that is at $\beta_2$ 
		can go at $\beta_1$.
	
		\item And so on, to the last couple of configurations: the 
		configuration that is at $\beta_{N_{(\beta)}}$ can go to 
		$\beta_{N_{(\beta)}-1}$ and that one that is at 
		$\beta_{N_{(\beta)}-1}$ can go at $\beta_{N_{(\beta)}}$ .

	\end{itemize}
	
\end{enumerate}

I stress the fact that you always try to swap {\em adjacent} 
$\beta$-values (or the procedure would be not effective). At 
a given point in time configuration number $27$ can be at 
$\beta_0$ and configuration number $3$ at $\beta_1$: these are 
the two configurations you will try to swap.

The $\beta$-Metropolis swap: as usual, you will accept if the 
swap makes the energy decreasing, and will accept with 
probability $e^{-\Delta S}$ if it makes the energy increasing.
You will have to compute

\bea
  \Delta S &=& S' - S \\
           &=&    ( \beta_{\alpha+1} E_{C_{\beta_\alpha}} +
                    \beta_{\alpha} E_{C_{\beta_{\alpha+1}}} ) -
                  ( \beta_{\alpha} E_{C_{\beta_\alpha}} +
                    \beta_{\alpha+1} 
                    E_{C_{\beta_{\alpha+1}}} )\ ,
           \nonumber 
\eea
where obviously $E_{C_{\beta_\alpha}}$ and 
$E_{C_{\beta_{\alpha+1}}}$ do not change during the 
$\beta$-swap. 

Here there is no freedom (and no need) for an additive term 
like the one we had before, $g_\alpha$.  In this method once 
fixed the $\beta$ values (the fixing of the $\beta$ can be 
done loosely, and does not need fine tuning) there are no free 
parameters, and no fine tuning needed. The ensemble of the 
parallel tempering is very different from the one of 
tempering, even if the two methods look very similar.

A possible way to look at the fact that we do not need the 
$g$ parameters is the following. The $g_\alpha$ were needed 
in order to prevent the system from collapsing to the state 
with lowest energy. But here there is a fixed number of 
$\beta$-values. A given value of $\beta_\alpha$, for example 
$\beta_{23}$, cannot disappear. This fact makes things easy 
for us.

\subsection{The Thermalization\protect\label{SS-THERMA}}

Thermalization is a crucial issue.  This is already true for 
usual spin models and usual local dynamics.  We want to be 
sure that we are looking at a system that has reached 
equilibrium, and after that we want to be sure we have 
correlation times under control.  In other words we are 
interested in studying the asymptotic equilibrium probability 
distribution, and we need to be sure that we are collecting 
the right information.  In complex dynamics like tempering, 
involving multiple Markov chains, it is even more difficult to 
keep correlation times under control.  The fact that we are 
typically studying complex systems, where slow dynamics is one 
of the most prominent feature, does not help.  I will discuss 
here some details of the understanding we have reached about 
this issue.  We will try to understand which kind of 
thermalization criteria we can adopt when using tempering.

\begin{figure}
  \epsfxsize=400pt\epsffile[28 30 566 320]{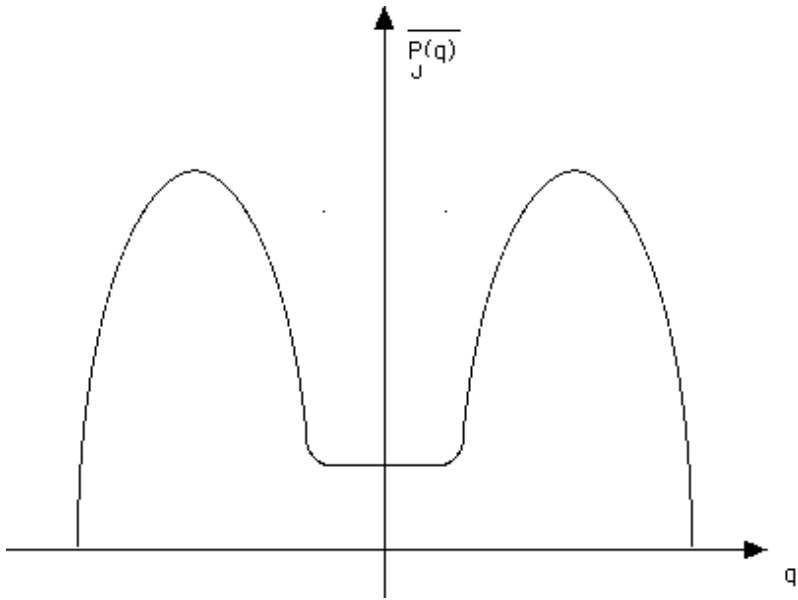}
  \caption[1]{
    $\overline{P_J(q)}$ versus $q$. The flat part close to 
    $q=0$ is the most remarkable feature of the behavior of a 
    class of disordered systems.
    }
  \protect\label{F-FIGTH1}
\end{figure}

Let us start by describing what happens when we use a simple 
Metropolis dynamics for studying a system with quenched 
disorder. Here we will not give details, that can be found for 
example in \cite{MEPAVI}, but I will only remind the reader 
that in this case one mainly deals with functions of the {\em 
overlap}

\be
  q^{\alpha\beta} \equiv \frac{1}{V} \sum_i
  \left(\sigma_i^\alpha \sigma_i^\beta\right)\ ,
\ee
where $\alpha$ and $\beta$ label two replica's of the systems, 
i.e.  two equilibrium configurations defined under the same 
realization of the quenched disorder.  The overlap gives us 
information about how similar are two typical equilibrium 
configurations of the same system.  $q$ plays the role of the 
order parameter, and it is the equivalent of the magnetization 
$m$ for usual spin models. The typical shape of the 
probability distribution of $q$, averaged over the quenched 
disorder $J$ (see \cite{MEPAVI}), is shown in fig. 
(\ref{F-FIGTH1}). The two peaks would be sharp (at $\pm 
m^{2}$) for an usual spin model.
The non-trivial part here is the fact that there 
is a continuous, non-zero contribution close to $q\simeq 0$: 
the system can exist at equilibrium in many states, that can 
even be, in the infinite volume limit, completely different.
The non-trivial part here is the fact that there 
is a continuous, non-zero contribution close to $q\simeq 0$: 
the system can exist at equilibrium in many states, that can 
even be, in the infinite volume limit, completely different.

In the case of a normal Metropolis update one can use two very 
strong criteria to check thermalization.

\begin{figure}
  \epsfxsize=400pt\epsffile[28 30 566 320]{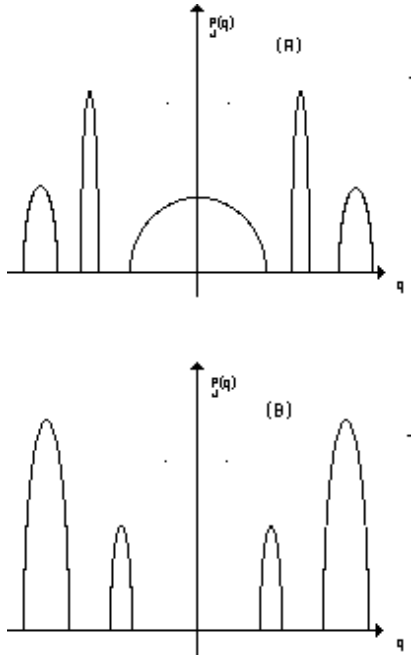}
  \caption[1]{
    Two typical $P_J(q)$ versus $q$, for two different 
    realizations of the coupling.
    }
  \protect\label{F-FIGTH2}
\end{figure}

\begin{enumerate}
	\item  {\em Check symmetry of $P(q)$ under $q\leftrightarrow 
	-q$ for each sample}. This is a very strong criterion. The 
	main issue is here that the full flip of the whole system, 
	$\{\sigma\}\to \{-\sigma\}$ is the slowest mode of the 
	dynamics. If you have done that well enough to get a good 
	$\pm \sigma$ symmetry you have explored the whole phase 
	space. In the simulation of a normal spin-model in the cold 
	phase we would never be so demanding (if not interested to 
	details of tunneling amplitudes): we would just seat in one 
	vacuum and compute observable quantities there. As we 
	already said, this is the main difficulty connected to the 
	study of disordered systems.
	
	In fig. (\ref{F-FIGTH2}) we give two typical $P(q)$ 
	for two different given quenched realizations of the 
	disorder. In the Parisi solution of the mean field (see 
	\cite{MEPAVI}) you can detailed compute many properties: for 
	example how many configurations contribute to the $q\simeq 
	0$ plateau. The numerical results in $3d$ show a very good 
	similarity with the mean field results 
	(\cite{MPRR,MAPARU}). 
	
	This criterion can be adopted for tempering, but in this case 
	it is not so strong anymore: in tempering methods the mode 
	$\{\sigma\}\to \{-\sigma\}$ is not necessarily a very slow 
	mode. It is possible in this case that one gets a very 
	symmetric $P(q)$ that is not related to the asymptotic 
	equilibrium distribution function.

	\item A second criterion, originally due to \cite{BHAYOU}, 
	that is very strong for usual dynamics, is based on using 
	two different definitions of the $P(q)$.
	\begin{enumerate}
	
		\item  After a random start from two different spin 
		configurations simulate two independent copies of the 
		system (in the same set of $\{J\}$ couplings). This is the 
		definition we had in mind till now. If we denote by $\sigma$ 
		and $\tau$ the two copies of the system we call $q_2$
		
		\be
		  q_2 \equiv \frac{1}{V}\sum_i(\sigma_i\tau_i)\ ,
		\ee
	     and $P_2(q)$ the equilibrium probability distribution of $q_2$.
	     
		\item In the second approach we use a dynamical measurement, 
		at different times. We wait for a large Monte Carlo time 
		separation $t$ and we define (for $t_0$ large enough)
		
		\be
		  q_{dyn} \equiv \frac{1}{V}\sum_i
		  \left(\sigma_i(t_0)\sigma_i(t_0+t)\right)\ ,
		\ee
		where eventually we will take an average over $t_0$. 
		We define $P_{dyn}(q)$ the equilibrium probability distribution of 
		$q_{dyn}$.
	\end{enumerate}
	
\end{enumerate}

When using the first definition at short times the two copies 
are not as similar as they will be asymptotically. 
Correlations only build up slowly. So $P_2(q)\to P(q)$ from 
below, and at short times $P_2(q)$ is centered at lower $q$ values 
than $P(q)$. On the contrary at short times $\sigma(t_0)$ and 
$\sigma(t_0+t)$ are correlated, i.e. $q_{dyn}$ at short times tends 
to be larger than the asymptotic value. So $P_{dyn}(q)\to P(q)$ from 
above. Only for times large enough  $P_{dyn}(q)=P_2(q)$, and we use 
this condition to check thermalization.

Also the extension of this second criterion to tempering is not 
straightforward, since now dynamics is assigning to different 
configurations different $\beta$ values.

For parallel tempering runs, for example, we use three main conditions 
to check thermalization.

\begin{enumerate}

	\item  We check that all observables are not drifting in time. We 
	look for example at the energy $E$ and at $q^2$. We check, on 
	logarithmic time scales, that they have safely converged to a stable 
	value. We also explicitly look at the full $P_J(q)$, and check it has 
	no drift.

	\item We check that the spacing of the allowed $\beta_\alpha$ is small 
	enough to guarantee a good acceptance ratio for the proposed $\beta$ 
	swaps.

	\item We demand that each of the $N_{(\beta)}$ configurations 
	$C_\alpha$ must have visited each of the allowed $\beta_\alpha$ 
	values with similar frequency. If we have done ten millions sweeps 
	and we have ten allowed $\beta$ values we demand that each 
	configurations has spend more or less one million sweeps in each 
	$\beta$ value. We keep a set of counters to check that. We want to 
	be able to detect situations where some systems are confined in a 
	part of the phase space, and a good acceptance factor is not enough 
	to avoid that (the configurations could just be flipping locally in 
	$\beta$ space).
	
\end{enumerate}

\subsection{More Comments\protect\label{SS-COMMEN}}

It is interesting to note some other technical comments about 
tempering like methods, mainly thinking about the large volume scaling 
behavior and about the performances of the method. Here we go, with a 
slightly miscellaneous series of comments.

When you increase the lattice size you need a larger number of 
allowed $\beta_\alpha$ values, i.e. a larger $N_{(\beta)}$, to sample 
the $\beta$-space, and to reach the region where free energy barriers 
disappear. The method is critically slowed down, but only like a 
power law.

Again, disordered systems need more attention than normal systems.  
You have to check that for each realization of the quenched disorder 
$\{J\}$ the condition that all $C_\alpha$ have visited all 
$\beta_\alpha$ values for similar periods of time is satisfied. From 
our experience the situation seems quite sharp: when the method works 
it works well, when it does not work it is a disaster (i.e. some 
visiting times are of order zero and some are of the order of the 
total time). For normal tempering (but parallel tempering seems 
preferable in all known cases) the constants $\{g_\alpha\}$ have to be 
tuned separately in each sample.

Let us discuss in better detail about volume scaling in tempering like 
methods. Let select the allowed $\beta$ values at a constant 
distance $\delta$:

\be
  \beta_{\alpha+1} = \beta_{\alpha} + \delta\ .
\ee
The probability for accepting a $\beta$-swap is

\be
   P_{SWAP}(\beta_{\alpha} , \beta_{\alpha+1}) \equiv  e^{-\Delta}\ ,
\ee
and

\be
  -\log(P_{SWAP}) = \Delta = 
  \delta \cdot 
  \left(
    S(C_{\beta_{\alpha+1}})-S(C_{\beta_{\alpha}}) 
  \right)
  \simeq \delta^2 \frac{dE}{d\beta}\ .
\ee
So if the specific heat is not diverging we want to select

\be
  \delta^2N=\mbox{constant}\ ,\ 
  \delta\simeq N^{-\frac12}\ ,\ \mbox{i.e.}\ 
  N_\beta \simeq  N^{\frac12}\ .
\ee
At a second order phase transition point, where the specific heat 
diverges, we have to be slightly more careful, since the number of 
intervals we need turns out to be higher. One has

\bea
  &\xi& \simeq |T-T_c|^{-\nu}\ ,\ C_V \simeq |T-T_c|^{-\alpha}\ , 
  \nonumber \\
  &|T-T_c|& \simeq \xi^{-\frac{1}{\nu}}\simeq N^{\frac{1}{d\nu}}\ ,
  \nonumber \\
  &C_V& \simeq  N^{\frac{\alpha}{d\nu}}\ .
\eea
So we get

\be
  \delta^2N^{1+\frac{\alpha}{d\nu}}=\mbox{constant}\ ,\ 
  N_\beta \simeq \delta^{-1} \simeq 
  N^{\frac12(1+\frac{\alpha}{d\nu})}\ ,
\ee
that is our final estimate.

The choice of the set $\{\beta_\alpha\}$ is not crucial (not like the 
$g_\alpha$'s in the serial tempering). We can select the same set for 
all the realizations of the disorder, paying only the price of loosing 
some small amount of efficiency.

\begin{figure}
  \epsfxsize=400pt\epsffile[28 30 566 350]{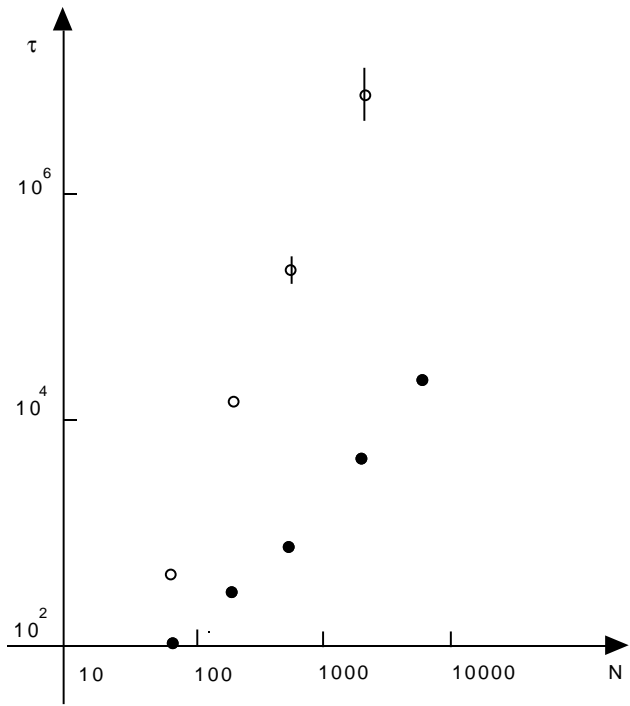}
  \caption[1]{
      Correlation times for multicanonical and for tempering, for 
      different lattice sizes (in log-log scale). The figure is only 
      sketchy, and the interested reader can find the exact data point 
      in fig. 2 the original paper by \protect\cite{HUKU2}. Filled 
      dots are for tempering, empty dots for multicanonical.
    }
  \protect\label{F-FIGCO1}
\end{figure}

In fig.  (\ref{F-FIGCO1}) we sketch (the picture only has a pictorial 
role) the correlation times computed by \cite{HUKU2}.  Filled dots are 
for tempering, empty dots for multicanonical.  Simulations are for a 
$3d$ Edwards Anderson spin glass, with couplings $J=\pm 1$.  $32$ 
$\beta$ values have been allowed in the parallel tempering run, for 
all $N$ values.  $\tau$ is defined as the typical time needed from a 
spin configuration for going from the cold to the warm phase. It is 
worth noticing that in the parallel tempering, given the $\beta$ set, 
there are no parameters to be fixed.

\subsection{Umbrella Sampling and Reweighting\protect\label{SS-UMBSAM}}

All the idea we have discussed up to now are related to the technique 
of Umbrella Sampling by \cite{TVA,TVB} (mainly developed for 
simulations of liquid systems).  Let us show now how we can relate 
that to reweighting techniques.

We consider an observable quantity $A$

\bea
\nonumber
\langle A\rangle_\beta &=&
\frac
{\int\{dC\}e^{-\beta S(\{C\})}  A(\{C\}) }
{\int\{dC\}e^{-\beta S(\{C\})}}
\\
&=& \int\{dC\} \pi_\beta (\{C\})  A(\{C\})\ ,
\eea
where $\pi_\beta$ is the Boltzmann equilibrium distribution at 
$\beta$,

\be
  \pi_\beta = \frac{e^{-\beta S(\{C\})}}{Z_\beta}\ .
\ee
So, now, we want to improve things: maybe dynamics at $\beta$ is slow, 
or we have already data at $\tilde{\beta}$ and we want an added bonus, 
or maybe we want to measure something more fancy (see earlier in the 
text). So we write

\be
\langle A \rangle_\beta =
\frac
{\int\{dC\}e^{-\beta S(\{C\})}  A(\{C\}) 
\frac{\tilde{\pi}}{\tilde{\pi}}}
{\int\{dC\}e^{-\beta S(\{C\})}\frac{\tilde{\pi}}{\tilde{\pi}}}
\left(
\frac
{\int\{dC\}\tilde{\pi}}
{\int\{dC\}\tilde{\pi}}
\right)\ ,
\ee
where we have inserted a large number of ones. We find in this way that

\be
\langle A \rangle_\beta =
\frac
{\langle \frac{A e^{-\beta S}}{\tilde{\pi}} \rangle_{\tilde{\pi}}}
{\langle \frac{  e^{-\beta S}}{\tilde{\pi}} \rangle_{\tilde{\pi}}}\ ,
\ee
where the expectation values are taken now over the $\tilde{\pi}$ 
probability distribution. If we choose

\be
  \label{RAINING}
  \tilde{\pi} = \frac{e^{-\tilde{\beta} S}}{Z_{\tilde{\beta}}}
\ee
we find the usual reweighting, as we have already discussed. But now 
we can also use the most general {\em umbrella sampling} by 
\cite{TVA,TVB}. The relation we have just derived, (\ref{RAINING}), is 
value for a generic probability distribution $\tilde{\pi}$. 
$\tilde{\pi}$ does not need to have the form of a Boltzmann 
distribution at some value of $\beta$, it can be everything you like. 
This is umbrella sampling: open $\pi\to\tilde{\pi}$, as an umbrella, 
to cover all of the parameter space in the region of interest. For 
example you can take

\be
  \tilde{\pi}(\{C\})
  =
  \sum_\alpha \omega(\beta_\alpha)
  e^{-\beta_\alpha}S(\{C\})\ .
\ee
Selecting the $\omega$ such that you get an equal sampling of the 
$\beta$ points gives $\omega\simeq e^f$, i.e. the choice of tempering.

\subsection{Multicanonical Methods\protect\label{SS-MULTIC}}

A large amount of work has been done on the so called {\em 
multicanonical methods}, \cite{BE31,BE32,BE1,BE2}, that are very powerful in 
order to study discontinuous phase transitions.  We discuss here a 
simple example, \cite{BE32}, where the $2d$, $10$ states Potts model is 
analyzed. The results we describe, by \cite{BE32}, are for lattices up 
to $100^2$. The action has the form

\be
  S = \sum_{\langle i,j\rangle} 
  \delta_{s_i,s_j}, \ s_i=0,1,\ldots, 9\ .
\ee
Such a model undergoes a temperature driven strong first order phase 
transition. A hard problem is to compute the interfacial free energy 
between the disordered and the ten ordered states. 

\begin{figure}
  \epsfxsize=400pt\epsffile[28 30 566 200]{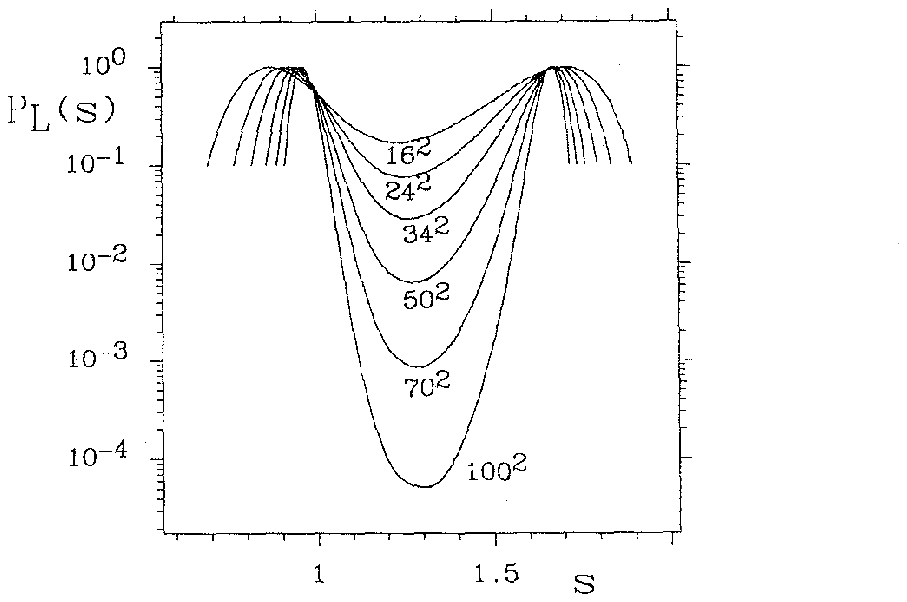}
  \caption[1]{
  The probability distribution of the energy for different 
lattice sizes.
    }
  \protect\label{F-FIGMC1}
\end{figure}

On a finite lattice there are no phase transitions: we define 
$\beta^c_L$, the pseudocritical coupling on a lattice of linear size 
$L$ such that the two peaks in the probability distribution of the 
internal energy have the same height. In fig. (\ref{F-FIGMC1}) (from 
\cite{BE32}) the probability  for different distribution of the energy
lattice sizes: on larger lattices the probability of getting a 
configuration in the forbidden region becomes smaller and smaller.
One has that

\begin{figure}
  \epsfxsize=400pt\epsffile[28 30 566 200]{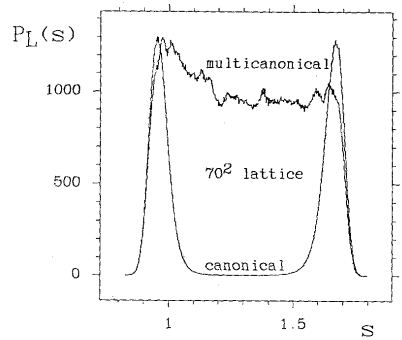}
  \caption[1]{
  The Boltzmann and the Multicanonical probability distribution 
  of the energy for $L=70$. The multicanonical distribution is quite 
  flat, allowing easy transitions among the two sides of the first 
  order phase transition.
    }
  \protect\label{F-FIGMC2}
\end{figure}

\be
  \label{PDIL}
  P_L^{(min)}\simeq e^{-\sigma L^{d-1}}\ .
\ee
When using a normal Metropolis dynamics eq. (\ref{PDIL}) implies that
the tunneling time diverges severely with the lattice size:

\be
  \tau_{Metropolis} \simeq AL^ae^{\sigma L^{d-1}}\ .
\ee
Multicanonical method takes a different approach, by modifying the 
sampling probability distribution. Here one samples phase space with 
weight

\be
  P_{L}^{MCan} \simeq e^{a_L^k - \beta_L^k S} \ , \ 
  \mbox{for} S_L^k < S \le S_L^{k+1}
\ee
(instead of usual $P_{L}^{Bolt} \simeq e^{-\beta_L S}$). One has 
partitioned the action range in intervals $I_k$, using a different 
action in each range. Now one chooses intervals $I_k$ and parameters 
$a_L^k$, $b_L^k$ such that $P_l^{(MCan)}$ is flat: fig. 
(\ref{F-FIGMC2}) shows that it can be done very successfully.

Configurations that were exponentially suppressed are now enhanced by 
the Multicanonical action: the interested reader can look at details 
about fixing the parameters in the original papers, \cite{BE31,BE32,BE1,BE2}.

In the case of the Multicanonical algorithm after data collecting you 
need reweighting (as opposed to tempering, where the output data at 
each $\beta$ value are directed distributed according to Boltzmann) to 
reconstruct the Boltzmann distribution. At $\beta^c_L$

\be
P_{L}^{Bolt} = 
e^{\beta^c_L S - \beta^k_L - a_L^k}
 P_{L}^{MCan}\ .
\ee
The improvement is dramatic, and an exponential slowing down becomes 
power like (\cite{BE32}: the same happens for tempering, for example in 
the cold phase plus to minus tunneling in the Ising model, \cite{FE}).
An estimate of \cite{BE32} gives $\tau_L^{MCan}\simeq 0.7L^{2.7}$, 
versus $\tau_L^{Heat Bath}\simeq 1.5L^{2.15}L^{0.08L}$.


\end{document}